\documentclass[pre,aps,amsmath,amsfonts,amssymb,showpacs,twocolumn,10pt]{revtex4-1}
\usepackage{epsfig,CJK}
\renewcommand{\emptyset}{\text{\large $\varnothing$}}
\begin{document}
\begin{CJK*}{UTF8}{mj}
\title{Critical decay exponent of the pair contact process with diffusion}
\author{Su-Chan Park (박수찬)}
\affiliation{Department of Physics, The Catholic University of Korea, Bucheon 420-743, Republic of Korea}
\date{\today}
\begin{abstract}
We investigate the one-dimensional pair contact process with diffusion (PCPD) 
by extensive Monte Carlo simulations, mainly focusing on the critical density decay exponent $\delta$.
To obtain an accurate estimate of $\delta$, we first find the strength of corrections to scaling 
using the recently introduced method [S.-C. Park. J. Korean Phys. Soc. {\bf 62}, 469 (2013)].
For small diffusion rate ($d\le 0.5$), the leading corrections-to-scaling term is found to 
be $\sim t^{-0.15}$, whereas for large diffusion rate ($d=0.95$) 
it is found to be $\sim t^{-0.5}$. After finding the strength of corrections to scaling, effective exponents
are systematically analyzed to conclude that the value of critical decay exponent $\delta$ is 
$0.173(3)$ irrespective of $d$.
This value should be compared with the critical decay exponent of the directed percolation, 0.1595. 
In addition, we study two types of crossover. At $d=0$, the phase boundary is
discontinuous and the crossover from the pair contact process
to the PCPD is found to be described by the crossover exponent $\phi = 2.6(1)$. 
We claim that the discontinuity of the phase boundary cannot be consistent 
with the theoretical argument supporting the hypothesis that the PCPD should
belong to the DP.
At $d=1$,
the crossover from the mean field PCPD to the PCPD is described by $\phi = 2$
which is argued to be exact.
\end{abstract}
\pacs{05.70.Ln, 05.70.Jk, 64.60.Ht}
\maketitle
\end{CJK*}
\section{\label{Sec:intro}Introduction}
The pair contact process with diffusion (PCPD) is an interacting particle system with diffusion, 
pair annihilation ($2 A \rightarrow 0$), and  branching by pairs ($2 A \rightarrow 3 A$). 
The PCPD was introduced in 1982 by Grassberger~\cite{G1982}, but had remained unnoticed in the 
statistical physics community for about 15 years. Since Howard and T{\"a}uber~\cite{HT1997} 
(re)introduced the so-called `bosonic' PCPD in 1997, the PCPD has been captivating statistical 
physicists and the effort to understand the critical behavior of the PCPD has continued 
until now~\cite{O2000,CHS2001,H2001,PK2002,BC2003,KC2003,JvWDT2004,NP2004,PHP2005a,PHP2005b,dOD2006,PP2006,KK2007,SB2008,PP2009,SB2012,GCDD2014}. 

It was accepted almost without question that the upper critical dimension of the PCPD is 2 and the 
PCPD does not belong to the directed percolation (DP) universality class in higher 
dimensions~\cite{OMS2002}. On the other hand, the one dimensional PCPD has gained its notorious fame 
from the beginning because of its strong corrections to scaling. Consequently, numerical studies 
have reported scattered values of critical exponents (see Table I of Ref.~\cite{SB2008} for a summary of reported values of critical exponents) and, in turn, as many hypotheses concerning the 
critical behavior were suggested as the number of research groups involved (for a review of the 
early various scenarios, see Ref.~\cite{HH2004}). It still remains an open question whether the one 
dimensional PCPD belongs to the DP class (DP hypothesis) or forms a different universality class 
just like the higher dimensional PCPD (non-DP hypothesis). Since we are mainly interested in the 
PCPD in one dimension, by the PCPD in the following we will exclusively mean the one-dimensional
PCPD, unless dimensions are explicitly stated.

To support the DP hypothesis, Hinrichsen~\cite{H2006} provided a theoretical argument as to why
the PCPD should belong to the DP class. This argument 
begins with the numerical observation that the dynamic exponent of the PCPD is smaller than 2, 
which means critical clusters spread super-diffusively. Since isolated particles can spread at most 
diffusively with dynamic exponent 2, diffusive motion of isolated particles can be regarded as 
frozen in comparison to the critical spreading and, in turn, the effectively frozen isolated particles 
can at best play the role of isolated particles of the pair contact process (PCP) without diffusion 
which is known to belong to the DP class. Accordingly, the PCPD should belong to the DP class.
Although this argument is not unquestionable (see Ref.~\cite{PP2008EPJB} for a critique), it is quite 
persuasive once the dynamic exponent of the PCPD is accepted to be smaller than 2 as numerical 
studies suggest. We will discuss this argument at the end of Sec.~\ref{Sec:pcp}.
Barkema and his colleagues have reported numerical results to support the 
DP hypothesis~\cite{BC2003,SB2008,SB2012}. 

At the same time, numerical studies supporting the non-DP hypothesis are also available. 
The critical behavior of the driven pair contact process with 
diffusion~\cite{PHP2005a,PHP2005b} seems to suggest that the PCPD 
cannot be described by a field theory with a single component field, 
which makes the PCPD not satisfy the prerequisite of the DP conjecture~\cite{J1981,G1982}.
Besides, the existence of non-trivial crossover from the PCPD
to the DP~\cite{PP2006,PP2009} was invoked to support the non-DP nature of the PCPD. 

Since the controversy arises from numerical difficulty of finding accurate value of critical exponents 
due to strong corrections to scaling, it is necessary to tame the corrections to scaling at our 
disposal. To this end, this paper exploits a systematic method suggested 
in Ref.~\cite{P2013} to find corrections to scaling without prior 
knowledge of leading asymptotic behavior. 
Once the strength of corrections to scaling is determined, the effective exponent
can be systematically analyzed to get an accurate estimate of the critical
exponent. In this paper, we find the critical decay exponent $\delta$
by extensive Monte Carlo simulations, using the method briefly mentioned above.

The paper is organized as follows: 
Section~\ref{Sec:model} consists of two parts. To be self-contained, 
Sec.~\ref{Sec:rule} introduces the dynamic rules of the PCPD and
describes the expected behavior of the order parameter in each phase. Also an algorithm to simulate
the stochastic dynamics is detailed with comparison to previous studies.
In Sec.~\ref{Sec:method}, a method to estimate the leading corrections-to-scaling term is explained.
Numerical estimate of the critical decay exponent is presented in Sec.~\ref{Sec:results},
using the method explained in Sec.~\ref{Sec:method}.
Section~\ref{Sec:cross} studies crossover behavior from
two extreme points of the model, $d=0$ (without diffusion) and $d\rightarrow 1$ with
a suitable time rescale (mean field), to the PCPD.
Section~\ref{Sec:sum} summarizes the work.

\section{\label{Sec:model}Model and Method}
\subsection{\label{Sec:rule}Model}
The PCPD is defined on a one-dimensional lattice of size
$L$ with periodic boundary conditions. Each site is either occupied by a 
particle ($A$) or empty ($\emptyset$) and every site can accommodate at most
one particle.  
The dynamics of the PCPD are defined by the following transition events,
\begin{eqnarray}
A \emptyset \leftrightarrow \emptyset A, &&\text{ with rate } d,\nonumber \\
AA \rightarrow \emptyset \emptyset, &&\text{ with rate } p (1-d),\nonumber\\
\emptyset AA, AA \emptyset \rightarrow AAA,&&\text{ with rate } (1-p)(1-d)/2,
\label{Eq:defmodel}
\end{eqnarray}
where $0 \le d < 1$ and $0 \le p \le 1$.
For bookkeeping purposes, we introduce a stochastic process $A_n(t)$ at every site $n$
which takes 1 (0) if site $n$ is occupied (vacant) at time $t$. 
We define the particle density, $\rho(t)$, the pair density, $\rho_p(t)$, and the triplet
density, $\rho_t(t)$, as
\begin{align}
\rho(t) &= \lim_{L\to\infty}\frac{1}{L}\sum_{n=1}^L \langle A_n(t) \rangle,\nonumber\\
\rho_p(t) &= \lim_{L\to\infty}\frac{1}{L}\sum_{n=1}^L \langle A_n(t) A_{n+1}(t) \rangle,\nonumber \\
\rho_t(t) &= \lim_{L\to\infty}\frac{1}{L}\sum_{n=1}^L \langle A_n(t) A_{n+1}(t) A_{n+2}(t)\rangle,
\end{align}
where $\langle \ldots \rangle$ means the average over ensembles. 
The evolution equation of $\rho(t)$ is
\begin{equation}
\label{Eq:EE}
\frac{1}{1-d}\frac{d \rho(t)}{dt} = (1 - 3 p) \rho_p(t)  - (1-p) \rho_t(t).
\end{equation}

When $p=p_c$ (critical point), both $\rho(t)$ and $\rho_p(t)$ are expected to show asymptotic 
power-law behavior with corrections to scaling as
\begin{align}
\label{Eq:rhoCTS}
\rho(t) &\sim a t^{-\delta} \left ( 1 + c t^{-\chi} + o(t^{-\chi}) \right ),\\
\rho_p(t) &\sim a't^{-\delta'} \left ( 1 + c' t^{-\chi'} + o(t^{-\chi'}) \right ),
\label{Eq:pair_asym}
\end{align}
where $a$, $a'$, $c$, $c'$ are constants
and $o(t^{-\chi})$ and $o(t^{-\chi'})$ contain higher
order terms which decay faster than $t^{-\chi}$ and $t^{-\chi'}$, respectively.  
In one dimension, it is believed that $\delta'$ equals $\delta$,
whereas the mean field theory assumes $\delta'=2 \delta$; see
Sec.~\ref{Sec:cross}. Also it is believed that $\chi = \chi'$, numerical
evidence of which will be provided in Sec.~\ref{Sec:results}. On this account, we will drop the
primes in the symbols of exponents for $\rho_p(t)$ in what follows and
we will refer to $\delta$ and $\chi$ as the critical decay exponent and
 the leading corrections-to-scaling exponent (LCSE), respectively.
In the active phase ($p<p_c$ within our model definition), both $\rho(t)$ and $\rho_p(t)$ approach
certain nonzero values exponentially as $t \to \infty$ and in the absorbing phase ($p>p_c$), 
$\rho(t)$ and $\rho_p(t)$ decrease to zero faster than $t^{-\delta}$ for nonzero $d$.

There are two important limiting cases. When $d=0$, this model
corresponds to the PCP~\cite{J1993} which has
infinitely many absorbing states and belongs
to the DP class. Meanwhile, taking $d \rightarrow 1$ limit with $\tau \equiv (1-d) t$ kept finite,
the (site) mean field theory becomes exact.
Hence there are two crossover behaviors at $d=0$ (from the PCP to
the PCPD) and at $d=1$ (from the mean field PCPD to the PCPD).
In Sec.~\ref{Sec:cross}, we will study these two kinds of crossover behavior.

To simulate the model, we employ the following algorithm. At first, we introduce 
\begin{equation}
dt \equiv 1/\text{max}(2d,1-d)
\end{equation}
which makes $2d\,dt$ and $(1-d)\,dt$ interpreted as probability. 
Now assume that there are $N(t)$ particles at time $t$. 
We randomly choose one particle among $N(t)$ particles with equal probability and
choose one of two nearest neighbors of the chosen particle with equal probability. Let us assume that
the chosen particle is located at site $n$ and the selected neighbor site is $n+k$ ($k = \pm 1$). 
If $A_{n+k}(t)=0$, the particle at site $n$ moves to site $n+k$ with probability $2 d\,dt$, but
with probability $1-2 d\,dt$, nothing happens. In the case $A_{n+k}(t)=1$,
two particles at sites $n$ and $n+k$ are removed with
probability $p(1-d)\,dt$ (pair annihilation), 
the site $n+2k$ becomes occupied with probability $(1-p)(1-d)\,dt$ (branching), 
or with probability $1 - (1-d)dt$ nothing happens.  
If $A_{n+2k}(t)$ is already 1 in the branching attempt, no change in
the configuration happens.
After the above attempt, time increases by $dt/N(t)$. 

Notice that the PCPD studied in Refs.~\cite{PHP2005a,PP2006} corresponds to the case with $d = \frac{1}{3}$
up to a time-rescale factor (time $t$ of the PCPD in Refs.~\cite{PHP2005a,PP2006} corresponds
to $\frac{3}{2} t$ of the case with $d=\frac{1}{3}$ in this paper).
Also note that the simulation algorithm employed in Ref.~\cite{O2003}
is slightly different from that used in this paper. But if we set 
$d = D/(2-D)$ where $D$ is the diffusion parameter used for simulations in Ref.~\cite{O2003}
and if we rescale time appropriately, the simulation results in Ref.~\cite{O2003} can be
directly compared to those obtained by the algorithm in the above.
For example, the reported critical point $\approx 0.133~53$ of the case with $D=0.5$ 
in Ref.~\cite{O2003} is consistent with that in Ref.~\cite{PP2006} which is $\approx 0.133~519$.

\subsection{\label{Sec:method}Corrections to scaling}
A systematic way to find the critical decay exponent simultaneously together with the critical point 
is to investigate the behavior of the effective exponent $-\delta_\text{eff}(t)$ defined as ($b>1$)
\begin{equation}
-\delta_\text{eff}(t) \equiv \frac{\ln \left [ \rho(t)/\rho(t/b)\right ]}{\ln b},
\end{equation}
which, by Eq.~\eqref{Eq:rhoCTS}, is expected to behave at the critical point as
\begin{equation}
-\delta_\text{eff}(t) = -\delta - c \frac{b^\chi -1}{\ln b}
t^{-\chi} + o(t^{-\chi}).
\label{Eq:rhoeff}
\end{equation}
In the time regime where higher order terms $o(t^{-\chi})$ are negligibly small
but the leading correction term $t^{-\chi}$ is not negligible, a plot of the effective 
exponent against $t^{-\chi}$ with the correct value of $\chi$ 
should be a straight line if the system is at the critical point. Meanwhile, it veers
up (down) if the system is in the active (absorbing) phase as $t^{-\chi}$ gets 
smaller. From the expected behavior in each phase, the critical exponent and the critical point can be 
found simultaneously by investigating the behavior of effective exponents near criticality, 
once the LCSE $\chi$ is known. Hence the information about the LCSE is indispensable in order
to estimate the critical decay exponent and the critical point accurately via the effective exponents.

A systematic method to find corrections to scaling without knowing $\delta$
was recently suggested~\cite{P2013}.
The idea is that at the critical point the double ratio of
$\rho$ at three different time points should behave asymptotically as
\begin{equation}
\left . \frac{\rho(t)}{\rho(t/b_1)} \right / \frac{\rho(t/b_1)}{\rho(t/b_1^2)} 
=1+ c (b_1^{\chi}-1)^2 t^{-\chi} + o(t^{-\chi}),
\label{Eq:oldCTSmeasure}
\end{equation}
where $b_1$ is a (fixed) constant.
Although $b_1$ is not necessarily the same as $b$ in Eq.~\eqref{Eq:rhoeff},
we will use the same value for $b$ and $b_1$ in this paper and we will drop 
the subscript in $b_1$ in the following. We introduce a corrections-to-scaling function (CTSF) $\Theta(t)$ 
as the logarithm of the left hand side of Eq.~\eqref{Eq:oldCTSmeasure}, 
\begin{align}
\Theta(t) = \ln \rho(t) + \ln \rho(t/b^2) -2 \ln \rho(t/b),
\end{align}
which behaves at criticality as
\begin{equation}
\Theta(t) \sim c (b^\chi-1)^2 t^{-\chi} + o(t^{-\chi}).
\label{Eq:AsymQ}
\end{equation}

\begin{figure}[t]
\includegraphics[width=\columnwidth]{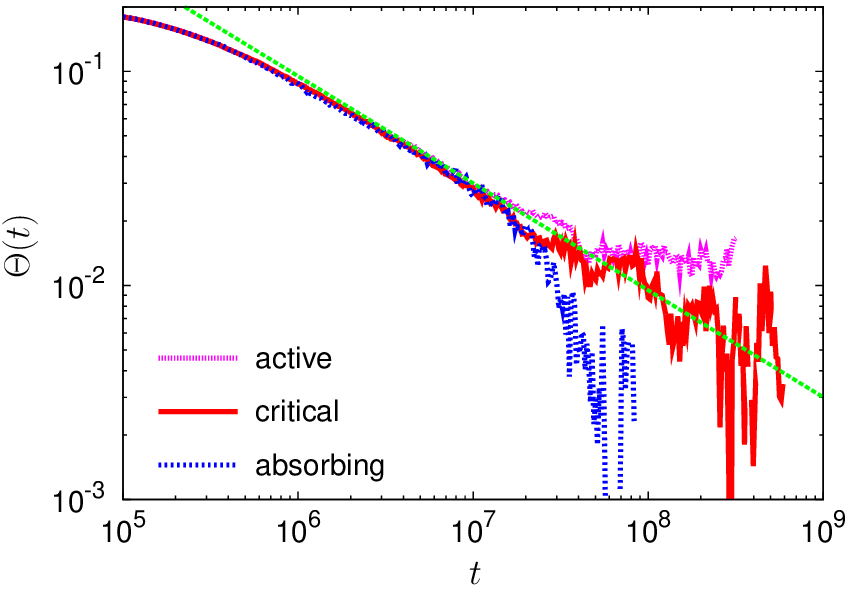}
\caption{\label{Fig:Qt} (Color online) Double logarithmic plots of $\Theta(t)$ vs $t$ 
around the critical point of the PCPD with $d=0.95$. The curves
correspond to $p=0.258~110$ (active), 0.258~112 (critical), and
0.258~114 (absorbing) from top to bottom; see Sec~\ref{Sec:results}.
The straight line with slope $-0.5$ is for guides to the eyes.}
\end{figure}

The behavior of $\Theta(t)$ at off-criticality is also of interest.
If the system is slightly away from the critical point with
$p = p_c + \Delta p$, $\Theta(t)$ is indistinguishable from Eq.~\eqref{Eq:AsymQ}
up to $t \sim | \Delta p |^{-\nu_\|}$, where $\nu_\|$ is the critical exponent
describing the divergence of correlation time.
Since $\Theta(t)$ can be understood as a curvature of
the curve $\ln \rho(t)$ as a function of $\ln t$, that is,
$\Theta(t) \approx \frac{d^2 \ln \rho(t)}{d \ln t ^2}$, $\Theta(t)$ gets 
larger (smaller) with $t$ if the system is in the active (absorbing) phase.
Hence, if the coefficient of $c$ in Eq.~\eqref{Eq:AsymQ} is positive, a typical behavior 
of $\Theta(t)$ around the critical point looks like Fig.~\ref{Fig:Qt}. 
These curves are obtained from simulations of the PCPD with $d=0.95$. To be specific,
the system size is $L=2^{21}$ and the
number of independent runs are 9800, 7200, and 1600 for $p=0.258~110$ (active), 0.258~112 (critical),
and 0.258~114 (absorbing), respectively.
Due to the double derivative-like nature of $\Theta(t)$, the curves obtained from 
numerical simulations can be very noisy as seen in Fig.~\ref{Fig:Qt} 
unless the number of independent simulation runs is very large.
In this respect, $\Theta(t)$ alone may not be a good measure to find the critical point and
the analysis of the effective exponents which are less noisy than $\Theta(t)$ should be accompanied.

Although the behavior of $\Theta(t)$ looks qualitatively similar to $\rho(t)$ around the critical point, 
$\Theta(t)$ in the active phase actually should approach zero as $t \rightarrow \infty$ because 
$\rho(t)$ in this limit saturates to a finite number with zero curvature. We will soon see such 
a long time behavior in the active phase from an exactly solvable model. 
In the absorbing phase, the behavior of $\Theta(t)$ in the limit of infinite time depends on the 
asymptotic behavior of $\rho(t)$. If $\rho(t)$ decreases exponentially, so does $\Theta(t)$. 
On the other hand, if $\rho(t)$ decreases as a power-law in the absorbing phase like the PCPD, 
$\Theta(t)$ should also approach zero as $t\rightarrow \infty$. 
Even though $\Theta(t)$ of the PCPD should approach zero in all phases as
$t\rightarrow \infty$, the deviation of $\Theta(t)$ at some point from the critical $\Theta(t)$ is 
conspicuous as Fig.~\ref{Fig:Qt} illustrates. Thus, such infinite time behavior does not limit the 
practical usefulness.

\begin{figure}[t]
\includegraphics[width=\columnwidth]{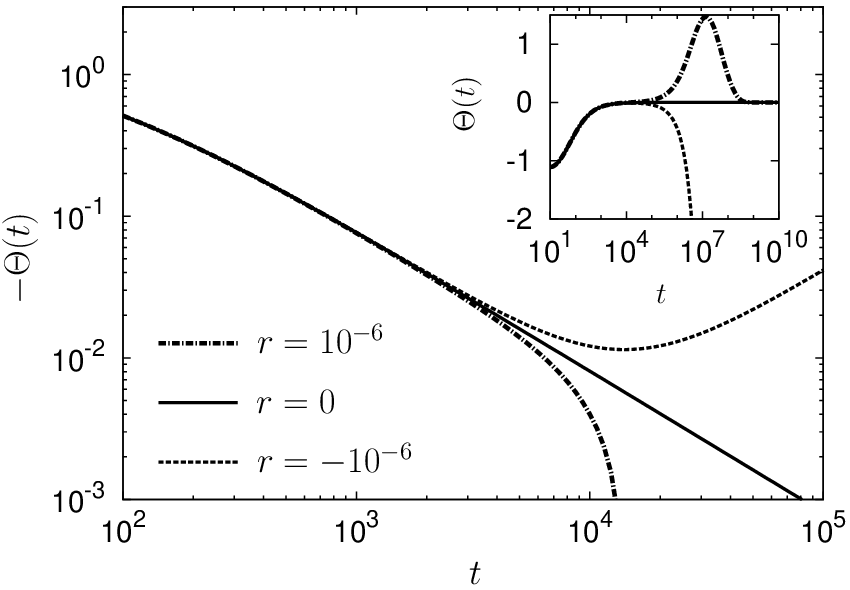}
\caption{\label{Fig:Qexnc} Log-log plots of $-\Theta(t)$ vs $t$ around the critical point 
of the model Eq.~\eqref{Eq:MFCP}. $\Theta(t)$ is calculated using Eq.~\eqref{Eq:MFCPsol} with $b=10$. 
Inset: Semi-logarithmic plots of $\Theta(t)$ vs $t$
for the same values of $r$'s as in the main figure. While $\Theta(t)$ for negative $r$ (in the absorbing phase) 
decreases exponentially, $\Theta(t)$
for positive $r$ (in the active phase) approaches 0 as $t\rightarrow \infty$.}
\end{figure}

The sign of $c$ is not necessarily positive and 
an example of the case with a negative $c$ can be found
by the following equation
\begin{equation}
\frac{d \rho(t)}{dt} = r \rho(t) - \rho(t)^2
\label{Eq:MFCP}
\end{equation}
with initial condition $\rho(t=0) = 1$. The solution is 
\begin{equation}
\rho(t) = \begin{cases}
(e^{-rt}+(1-e^{-rt})/r)^{-1}, & r\neq 0,\\
(1+t)^{-1} \sim t^{-1}\left (1-t^{-1}\right ), & r=0.
\end{cases}
\label{Eq:MFCPsol}
\end{equation}
Since the coefficient of the leading corrections-to-scaling term at the critical point ($r=0$) is negative,
it is appropriate to draw $-\Theta(t)$ vs $t$ on a double logarithmic scale as
in Fig.~\ref{Fig:Qexnc} which makes the curve in the active (absorbing) phase veer down (up) 
contrary to Fig.~\ref{Fig:Qt}. The inset of Fig.~\ref{Fig:Qexnc} shows that
$\Theta(t)$ in the active phase approach 0 as $t \rightarrow \infty$ as argued before.
Since $\rho(t)$ decreases exponentially in the absorbing phase of this example, 
$\Theta(t)$ also decreases exponentially to $-\infty$.

\section{\label{Sec:results}Critical Decay Exponent}
This section investigates the critical decay exponent $\delta$ of the
PCPD for various $d$'s via the analysis of the effective exponents 
along with the corresponding CTSFs. For all numerical analyses in this section, 
the system size is $L = 2^{21}$ and the initial density is 1. 
Since $\rho(t)$ and $\rho_p(t)$ are equally important, the CTSFs for
both $\rho(t)$ and $\rho_p(t)$ are studied and will be denoted by 
$\Theta_r(t)$ and $\Theta_p(t)$, respectively.

\begin{figure}[t]
\includegraphics[width=\columnwidth]{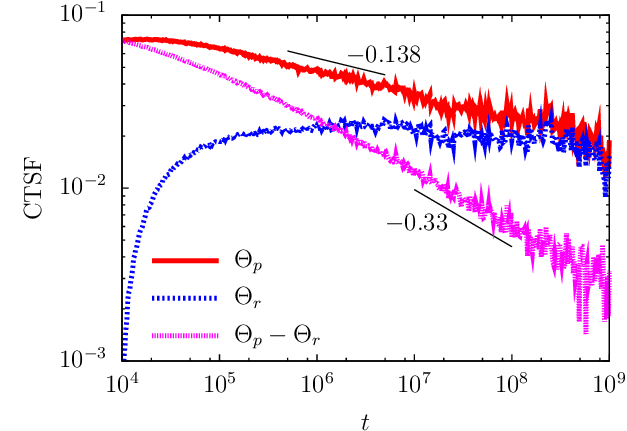}
\caption{\label{Fig:d01} (Color online) Log-log plots of CTSFs vs $t$ 
at $p=0.111~158$ and $d=0.1$. Here, $b=10$ is used. Two line segments with 
slopes $-0.138$ and $-0.33$ as indicated in the figure are also drawn for 
guides to the eyes. }
\end{figure}

At first, we will present the results for the case with $d=0.1$. As we will 
show later, the critical point is found to be $p_c = 0.111~158(1)$, where the 
number in parentheses indicates the uncertainty of the last digit. 
Figure~\ref{Fig:d01} shows double-logarithmic plots of
$\Theta_r$ and $\Theta_p$ vs $t$ at $p=0.111~158$ for $b= 10$. 
These data are obtained from 3000 independent runs at the designated value of $p$.
It seems that $\Theta_p$ shows a
power-law behavior as $\Theta_p \sim t^{-0.138}$ while $\Theta_r$
has no symptom of power-law decay up to $t = 10^{8}$. On the other hand,
$\Theta_r$ and $\Theta_p$ become almost overlapped after $t = 10^{8}$,
which implies that not only the LCSE but also the coefficients of the leading 
correction-to-scaling terms
of $\rho$ and $\rho_p$ are identical. 
Since $\Theta_p$ exhibits a more or less clean power-law behavior and
$\Theta_r$ eventually follows $\Theta_p$, we estimate $\chi$ to be 0.138 from $\Theta_p$. 
Note that this estimate is comparable to that in Ref.~\cite{SB2012}.

It is worthwhile to discuss the implication of the difference between the two CTSFs, 
$\Theta_p - \Theta_r$, which shows
a clean power-law behavior with $\simeq t^{-0.33}$ albeit $\Theta_p \sim t^{-0.138}$.
By definition of the CTSF, $\Theta_p - \Theta_r$ can be regarded
as the CTSF for $\rho_p(t)/\rho(t)$.
Hence if one analyzes $\rho_p(t)/\rho(t)$ rather than $\rho(t)$ and $\rho_p(t)$
separately with the assumption of $c\neq c'$ in Eqs.~\eqref{Eq:rhoCTS} and \eqref{Eq:pair_asym},
one may wrongfully conclude that the corrections to scaling 
of the PCPD are weaker than the actual strength $t^{-0.138}$. 
In fact, the cancellation of the leading corrections-to-scaling term
in $\rho_p(t)/\rho(t)$ was already anticipated in Ref.~\cite{SB2012} 
and we confirmed it through the direct numerical analysis.

Actually, the cancellation of the leading corrections-to-scaling term in the ratio
of two quantities is not unusual. An immediate example
arises when we analyze Eq.~\eqref{Eq:EE}. Since $\chi<1$, 
$\rho_p(t)$ and $\rho_t(t)$ at the critical point should be
\begin{align}
\rho_p(t) &= a_p t^{-\delta} \left ( 1 + 
c t^{-\chi} + \Xi(t) + e_p t^{-1} + o(t^{-1})\right ),\nonumber\\
\rho_t(t) &= a_t t^{-\delta} \left ( 1 + 
c t^{-\chi} + \Xi(t) + e_t t^{-1} + o(t^{-1})\right ),
\end{align}
where $a_p$ and $a_t$ are constants satisfying $a_t (1-p_c) = a_p (1-3p_c)$,
$\Xi(t)$ contains all terms decaying faster than $t^{-\chi}$ but slower than $t^{-1}$, and
$e_p$ should be strictly smaller than $e_t$.
Otherwise, the leading power of the left hand side of Eq.~\eqref{Eq:EE} cannot be the same
as that of the right hand side of the equation. Hence, $\rho_p(t)/\rho_t(t)$ at the 
critical point should have the form
\begin{equation}
\frac{\rho_p(t)}{\rho_t(t)} = \frac{1-p_c}{1-3 p_c} \left ( 1 + (e_p-e_t) t^{-1} + o(t^{-1}) \right ).
\end{equation}
Unlike the cancellation of the leading corrections-to-scaling terms in $\rho_p/\rho_t$,
however, we could not find any theoretical reason why $\rho(t)$ and $\rho_p(t)$ should have
exactly the same leading corrections-to-scaling term.
This can be a theoretical challenge of the PCPD.

\begin{figure}[t]
\includegraphics[width=\columnwidth]{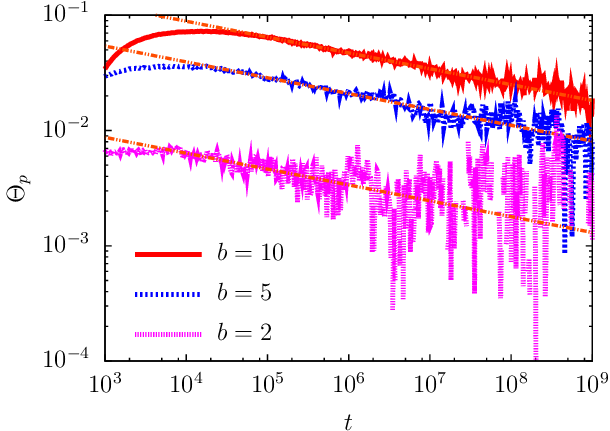}
\caption{\label{Fig:d01(b)} (Color online)  Plots of $\Theta_p$ vs $t$ 
at $p=0.111~158$ and $d=0.1$ for $b= 10$, $5$, and $2$, from top to bottom. 
The straight line lying on $\Theta_p$ with $b=10$ is the result of
the power-law fitting $c (b^{\chi}-1)^2 t^{-\chi}$ with
fitting parameters $c$ and $\chi$. 
Two straight lines lying on $\Theta_p$'s for $b=5$
and $b=2$, respectively, are plots of Eq.~\eqref{Eq:AsymQ} with
$c$ and $\chi$ obtained from the fitting.
}
\end{figure}
To check the consistency of Eq.~\eqref{Eq:AsymQ}, we analyze $\Theta_p$'s for
various values of $b$ ($b=5$ and $b=2$). We first fit $\Theta_p$ for $b=10$ 
using a fitting function $c (b^\chi-1)^2 t^{-\chi}$ with two fitting
parameters $c$ and $\chi$. From the fitting, we estimate $\chi \approx 0.138$ and the coefficient 
of the leading corrections-to-scaling term to be $c \approx 2.27$. The straight
line lying on $\Theta_p$ for $b=10$ (top curve) in Fig.~\ref{Fig:d01(b)} 
is the result of this fitting. Then, we compare $c (b^\chi-1)^2 t^{-\chi}$
for $b= 5$ and $b=2$ with the estimated values of $c$ and $\chi$ to
$\Theta_p$'s for corresponding $b$'s, which shows an excellent coincidence. 

We think this coincidence provides a numerical evidence for the absence of 
logarithmic corrections in the leading corrections-to-scaling term. When we 
derive Eq.~\eqref{Eq:AsymQ}, we tacitly assumed that the leading term has 
no logarithmic corrections. If there happens
to be such corrections, the above procedure should exhibit a systematic
deviation for different values of $b$.
Also a nice power-law behavior of $\Theta_p$ for about four decades, as
seen in Fig.~\ref{Fig:d01(b)} suggests that
logarithmic corrections, even if exist, are negligible. Furthermore,
the clean power-law behavior of $\Theta_p - \Theta_r$ also provides an
indirect evidence against logarithmic corrections. 

\begin{figure}[t]
\includegraphics[width=\columnwidth]{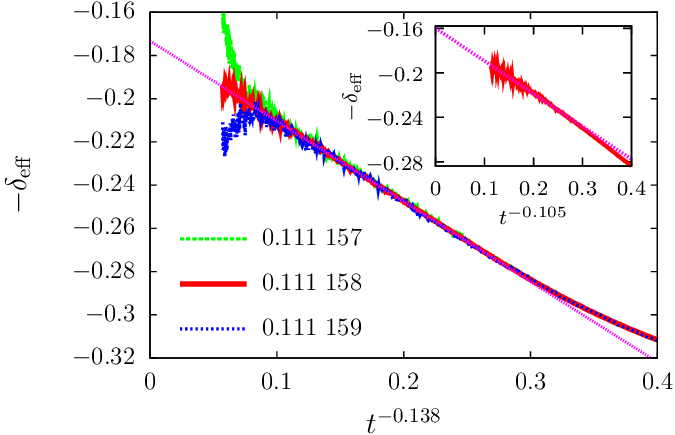}
\caption{\label{Fig:d01eff} (Color online) 
Plots of $-\delta_\text{eff}$ vs $t^{-0.138}$ 
for $p=0.111~157$, 0.111~158, and 0.111~159 with $d=0.1$, from top to bottom. 
$b$ is set 10.
The straight line which
intersects the ordinate at $\approx -0.173$ is a fitting result of $-\delta_\text{eff}$ for $p=0.111~158$. Inset: Plot of $-\delta_\text{eff}$ vs $t^{-0.105}$
at the critical point. The linear extrapolation gives the DP critical exponent.
}
\end{figure}

Finding the LCSE to be 0.138, we now analyze the effective exponent
$-\delta_\text{eff}$. 
Since $\Theta_p$ shows a cleaner power-law behavior than $\Theta_r$, 
we analyze $-\delta_\text{eff}$ calculated from $\rho_p(t)$.
In Fig.~\ref{Fig:d01eff}, $-\delta_\text{eff}$ obtained using $b=10$ 
is drawn as a function of $t^{-\chi}$ with $\chi = 0.138$ near criticality. 
From this plot,
it is clear that the system with $p=0.111~157$ (0.111~159) is in 
the active (absorbing) phase and the critical point should 
be $p_c = 0.111~158(1)$. Although the number of independent runs for both off-critical simulations 
is only 200, the effective exponents give a clear illustration of the expected behavior in each phase. 
We fit $-\delta_\text{eff}$ for $p=0.111~158$ using a linear function to obtain that $\delta\approx0.173$. Since a fitting result of $\delta$ varies from 
0.17 (for $\chi=0.13$)
to 0.176 (for $\chi = 0.15$) with $\chi$, we conclude that
$\delta = 0.173(3)$.

Figure~\ref{Fig:d01eff} shows that $-\delta_\text{eff}$ at criticality 
becomes an almost straight line in the region $t^{-0.138} \le 0.25$. 
This behavior is indeed consistent with the analysis of $\Theta_p$.
We observed in Fig.~\ref{Fig:d01(b)} that $\Theta_p$ for $b=10$ exhibits
a nice power-law behavior from $10^5$. Recall that $\Theta_p$ is calculated 
using $\rho_p$ at $t$, $t/b$, and $t/b^2$. Thus, this numerical observation
implies that the leading corrections-to-scaling
term becomes dominant from $t =  10^3$. Thus $-\delta_\text{eff}$ 
for $b=10$ should be almost straight from
$10^{-4 \chi} \approx 0.25$, as seen in Fig.~\ref{Fig:d01eff}. 

Since the estimated critical exponent 0.173 is close to the DP exponent 0.1595,
it would be an interesting practice which value of $\chi$ can predict the DP 
critical exponent. By trial and error, we found that $\chi = 0.105$ gives the
DP critical exponent; see the inset of Fig.~\ref{Fig:d01eff}.
Although 0.105 is different from the estimated 0.138 by 25\%, 
it is indeed hard to exclude the possibility of the DP critical
scaling. Thus, the analysis of the PCPD with $d=0.1$ alone may not
be enough to conclude that the PCPD does not belong to the DP class.
To make the estimate 0.173 more convincing, we analyze other cases with 
different values of $d$.

\begin{figure}[t]
\includegraphics[width=\columnwidth]{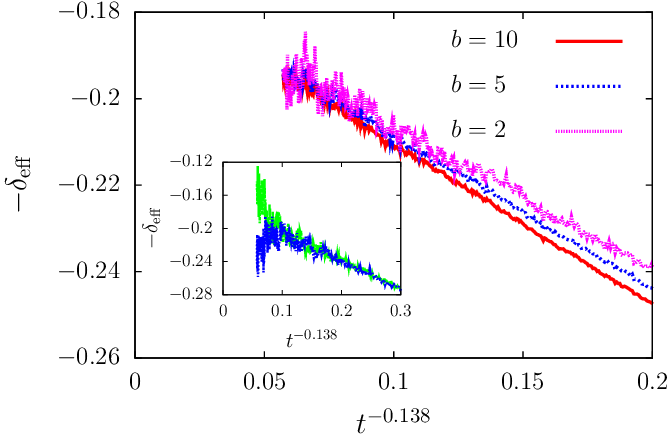}
\caption{\label{Fig:d01eff(b)} (Color online) 
Plots of $-\delta_\text{eff}$ vs $t$ at $p=0.111~158$ for
$b=10$, 5, and 2 (from bottom to top). 
Inset: $-\delta_\text{eff}$'s for $b=2$ are plotted against $t^{-0.138}$
at $p=0.111~157$ (top) and 0.111~159 (bottom).
}
\end{figure}

Before delving into the case with different diffusion probability,
we will show that the estimated critical point is rather insensitive
to the estimates of $\delta$ and $\chi$.
If $\chi$ is small as in the present case, 
we can approximate $(b^\chi -1)/\ln b \approx \chi
+ O(\chi^2 \ln b)$  as long as $\ln b$ is not so large.
Thus, the effective exponent at criticality should be  
insensitive to $b$.
As can be seen in Fig.~\ref{Fig:d01eff(b)}, $-\delta_\text{eff}$ at
$p=0.111~158$ is more or less insensitive to $b$, which strongly supports that 
the density at $p=0.111~158$ exhibits the
critical scaling up to the simulated time.
Meanwhile, if the system is in the active (absorbing) phase
and $t > |p-p_c|^{-\nu_\|}$, $-\delta_\text{eff}$ at given $t$ should
increase (decrease) significantly as $b$ gets smaller.
The inset of Fig.~\ref{Fig:d01eff(b)} depicts
$\delta_\text{eff}$ at $p=0.111~157$ and $0.111~159$ for $b=2$.
By comparing this figure with with Fig.~\ref{Fig:d01eff}, 
it is easily recognized that $\delta_\text{eff}$ at off-criticality is significantly affected 
by the change of $b$.
Although we plotted $-\delta_\text{eff}$ vs $t^{-0.138}$ in 
Fig~\ref{Fig:d01eff(b)} for convenience, 
different choice of $\chi$ does not affect the observed behavior 
of the effective exponents under the change of $b$.
Also note that $\delta$ does not play any role in the above discussion.
Hence, we conclude that the estimated critical point
$p_c = 0.111~158 (1)$ is accurate irrespective of whether we
are using the right value of $\delta$ and $\chi$.

\begin{figure}[t]
\includegraphics[width=\columnwidth]{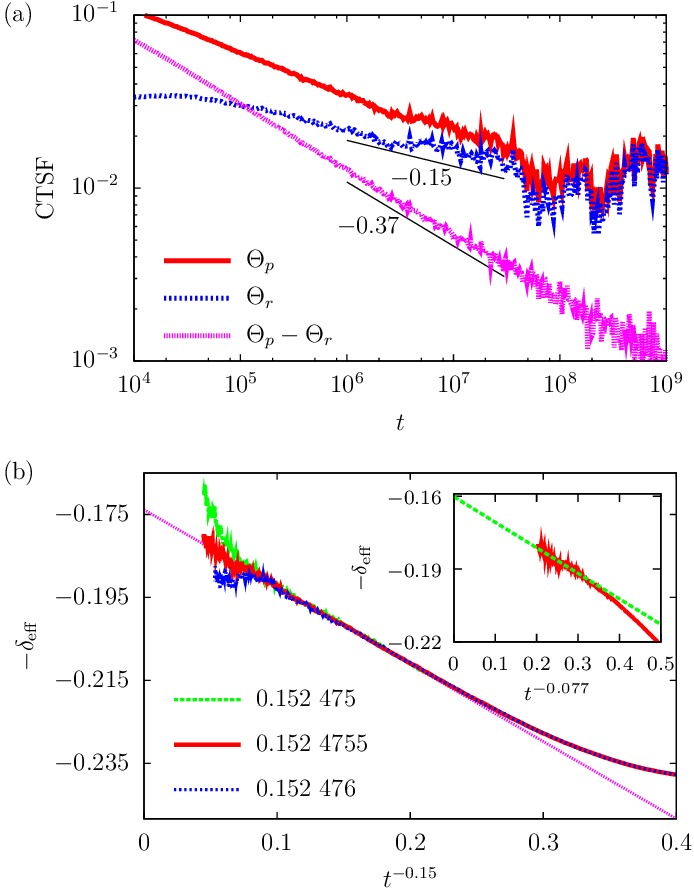}
\caption{\label{Fig:d05} (Color online) (a) Log-log plots of CTSFs vs $t$ 
at $p=0.152~4755$ and $d=0.5$. Two line segments with slopes
$-0.15$ and $-0.37$ as indicated in the figure are also drawn for guides
to the eyes. (b) Plots of $-\delta_\text{eff}$ vs $t$ 
for $p=0.152~475$, 0.152~4755, and 0.152~476 with $d=0.5$, from top to bottom. 
The straight line which
meets the ordinate at $\approx -0.174$ is a fitting result of the
data for $p=0.152~4755$. Inset: Plot of $-\delta_\text{eff}$ vs $t^{-0.077}$
at the critical point. The linear extrapolation gives the DP critical exponent.
}
\end{figure}

Now, we will analyze the case of $d=0.5$.
Figure~\ref{Fig:d05}(a) depicts the CTSFs as functions of $t$ on a double logarithmic scale at 
$p=0.152~4755$ which will turn out to be the critical point. The data are collected from
7000 independent simulation runs and $b=10$ is used.
Unlike the case with $d=0.1$, $\Theta_r$ shows a power-law decay $t^{-0.15}$
from about $t=10^5$. $\Theta_p$ in the short time regime decays
faster than $\Theta_r$ but after $t=10^8$, $\Theta_r$ and
$\Theta_p$ are almost indistinguishable. Since $\Theta_r$ shows a
more stable power-law behavior than $\Theta_p$, we estimate
$\chi$ to be 0.15 which is comparable to the above estimate. One can also
see that $\Theta_p - \Theta_r$ decays faster than $t^{-0.15}$.

In Fig.~\ref{Fig:d05}(b), $-\delta_\text{eff}$ corresponding to $\rho(t)$ for
$b=10$ is drawn against $t^{-0.15}$. 
The effective exponent for $p = 0.152~475$ (0.152~476) 
results from 2400 (2500) independent simulation runs. This figure shows that 
the critical point is
$p_c = 0.152~4755(5)$ and a linear fit of $-\delta_\text{eff}$ for
$p=p_c$ gives $\delta \approx 0.174$ which is consistent
with the estimate for the case of $d=0.1$. Also note that 
$-\delta_\text{eff}$ is
almost straight in the region $t^{-\chi} < 10^{-4\chi} \approx 0.25$, which
is consistent with the behavior of $\Theta_r$.

We would like to emphasize that the estimate of the critical point is 
rather insensitive to the accuracy of $\delta$ and
$\chi$, so the accuracy of $p_c$ is less questionable than
the exponents. Using the data of the density at the critical point,
we also estimate, by trial and error, the value of $\chi$ which 
gives the DP exponent. At this time, the desired value of 
$\chi$ becomes $0.077$ which is quite different from the estimated 
$\chi = 0.15$. Furthermore, 0.077 differs by 25\% from the case of
$d=0.1$. That is, for our numerical data 
to be consistent with the DP hypothesis, the LCSE has to vary
continuously with $d$ significantly.

Since the case of $d=0.5$ was also studied in Ref.~\cite{SB2012} which
supports the DP hypothesis, 
it is worth while to compare our results with those in Ref.~\cite{SB2012}.
First, the critical point in this paper is more accurately estimated
than in Ref.~\cite{SB2012}; see Table I of Ref.~\cite{SB2012}. Second,
a value close to the DP exponent was obtained from the system at $p=0.152~473$ which is actually in the active phase. Because the density in the active phase 
decays slower than at the critical point, it is not surprising that
the estimated value of $\delta$ in Ref.~\cite{SB2012} is
smaller than 0.173. Interestingly, however, a similar
estimate of $\delta$ was attained when the system at $p=0.152~476$ was analyzed;
see the sixth row of Table 1 in Ref.~\cite{SB2012}. 

\begin{figure}[t]
\includegraphics[width=\columnwidth]{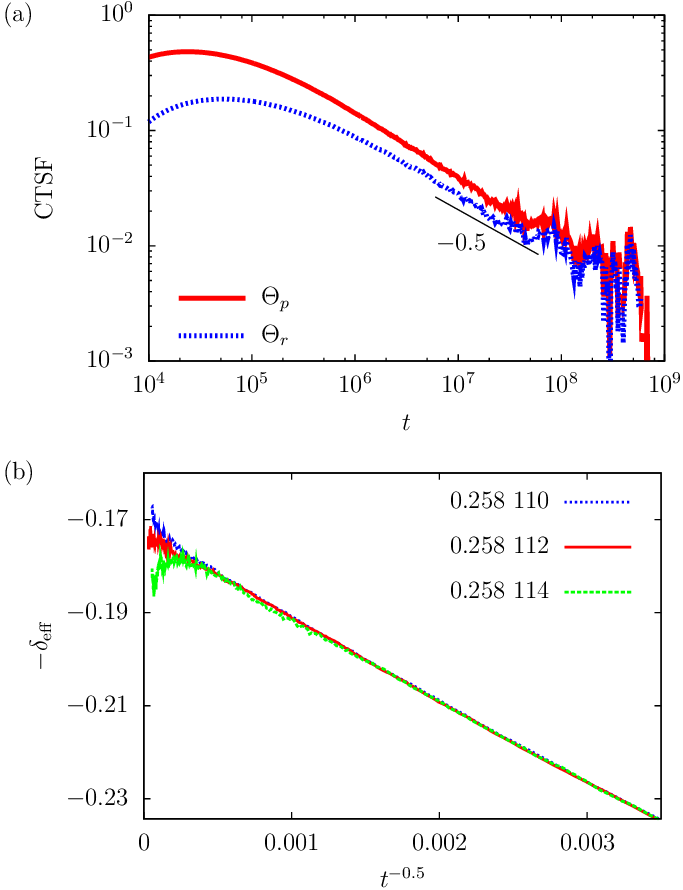}
\caption{\label{Fig:d95} (Color online) (a) Log-log plots of $\Theta_p$  
(top curve) and $\Theta_r$ (bottom curve) against $t$ at $p=0.258~112$ and $d=0.95$. 
A line segment with slope $-0.5$ is a guide to the eyes. (b) Plots of $-\delta_\text{eff}$ vs $t^{-0.5}$ 
for $p=0.258~110$, 0.258~112, and 0.258~114 with $d=0.95$, from top to bottom. 
}
\end{figure}
The results up to now seem to suggest that the LCSE obtained from
the behavior of CTSFs is about 0.15 for the PCPD in general. 
Quite intriguingly, however, the PCPD with $d = 0.95$ has relatively weak corrections to scaling.
In Fig.~\ref{Fig:d95}(a), the CTSFs for $b=10$ at the critical point of
the case with $d=0.95$ are depicted as functions of $t$ on a double-logarithmic
scale. Note that $\Theta_r$ in this figure is the middle curve in 
Fig.~\ref{Fig:Qt}. 
Unlike the previous two cases, $\Theta_r$ decays as $t^{-0.5}$ 
rather than $t^{-0.15}$. Note that the power-law behavior of $\Theta_r$ is
observed from $t=10^6$ which means the LCSE in $\rho(t)$ 
becomes dominant from $t=10^4$. 

Using this LCSE, we depict $-\delta_\text{eff}$ calculated from
$\rho(t)$ with $b=10$ as a function of $t^{-0.5}$ in Fig.~\ref{Fig:d95}(b) 
which again shows a typical behavior of $-\delta_\text{eff}$ near criticality.
From the behavior of $-\delta_\text{eff}$ we estimate $p_c = 0.258~112(2)$. 
A linear fit of
$-\delta_\text{eff}$ for $p = 0.258~112$ suggests $\delta \approx 0.173$,
which is again consistent with the estimates from the cases of different $d$'s.
Since the LCSE is dominant from $10^{4}$, the effective
exponent should be a straight line for $t^{-\chi} < 10^{-5\chi} \approx 0.003$,
as seen in Fig.~\ref{Fig:d95} (b).

We also investigated which value of $\chi$ can give the DP exponent
for the high diffusion case. Unlike
the low diffusion cases, however, the DP exponent was hardly observed
by varying $\chi$, which seems to imply that 
the critical behavior of the PCPD with $d=0.95$ cannot be consistent with the 
DP hypothesis. 

\begin{figure}[t]
\includegraphics[width=\columnwidth]{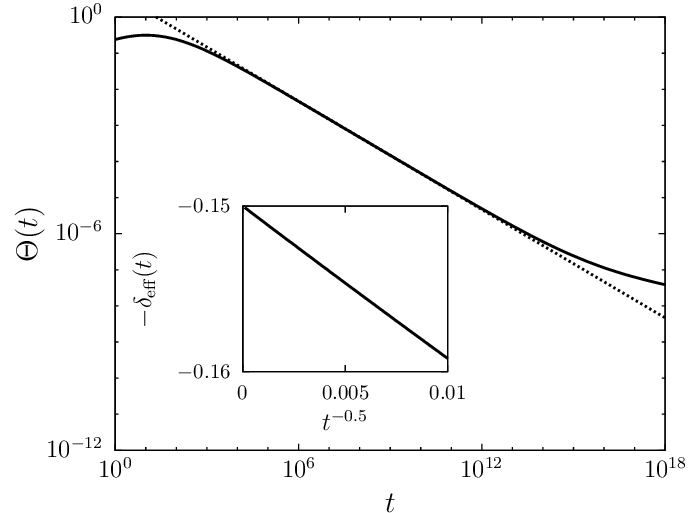}
\caption{\label{Fig:toy} Log-log plot of $\Theta$ vs $t$ of the toy equation. The straight
line with slope $-0.5$ is a guide to the eyes. Up to $t=10^{12}$,
$\Theta(t)$ shows a nice power-law behavior of $t^{-0.5}$. Inset: Plot of $-\delta_\text{eff}$ vs $t^{-0.5}$.
Although the true leading corrections-to-scaling term of Eq.~\eqref{Eq:toy} is 
$t^{-0.15}$, $-\delta_\text{eff}$ drawn as a function of $t^{-0.5}$ gives the exact leading
term.}
\end{figure}
Since $\chi$ needs not be universal, appearing dependence of $\chi$ on $d$ is 
not contradictory to our common sense formed by the renormalization group (RG) 
theory. 
Still, what kind of mathematical structure is behind the change of $\chi$ with $d$
is an interesting question. 

We think there are two possible scenarios. The first obvious scenario is 
that $\chi$ is a continuous function of $d$. Without a fixed line in the RG sense, however, it is 
hard to expect such a continuously varying exponent albeit non-universal, so this scenario does not look 
plausible.
The second one is that the correction term $t^{-0.15}$ is actually present even in the case with 
$d=0.95$ but the coefficient is very small. 

The implication of the second scenario can be clearly stated by
the following toy equation
\begin{equation}
\rho(t) = t^{-0.15} \left ( 1 + 10^{-4} t^{-0.15}  + t^{-0.5} \right ).
\label{Eq:toy}
\end{equation}
In this toy example, the leading corrections-to-scaling term becomes dominant only when 
$t \gg 2.6\times 10^{11}$ and before this time $t^{-0.5}$ plays the role of the leading 
corrections-to-scaling term.
Thus, $\Theta(t)$ is well described by $t^{-0.5}$ and the corresponding effective
exponent drawn as a function of $t^{-0.5}$ rather than the true asymptotic behavior $t^{-0.15}$ gives
the correct value of 0.15; see Fig.~\ref{Fig:toy}.
We think the second scenario is plausible, but more investigation is necessary to fully
understand how the mathematical structure of the corrections to scaling changes with $d$.
This also can be a theoretical challenge of the PCPD.
In any case, $\Theta(t)$ is a useful tool to find the correct value of the critical exponent, although it may not predict the true LCSE as in the toy
example.

To conclude this section, we found that the critical decay exponent of the 
PCPD is robust against $d$ with value $\delta= 0.173(3)$. 
Although this value differs from the DP value only by 8\%, 
the consistent estimate for a wide range of $d$ 
supports the non-DP hypothesis. In particular, the DP hypothesis 
is not consistent with the case with $d=0.95$ up to the simulation time
in this work ($t=10^9$).
Since the corrections-to-scaling for
the high diffusion case is much weaker than those for the low diffusion case, 
it seems promising to find other exponents accurately by investigating 
the PCPD for large $d$.

\begin{table}[t]
\caption{\label{Table:pc}Critical points of the PCPD for various $d$. The numbers in parentheses indicate uncertainty of the last digits.}
\begin{ruledtabular}
\begin{tabular}{ll}
$d$&$p_c$\\
\hline
0&0.077~0905(5)\footnotemark[1]\\
0.001&0.1019(1)\footnotemark[2]\\
0.005&0.1023(1)\footnotemark[2]\\
0.01&0.1028(1)\footnotemark[2]\\
0.02&0.1038(1)\footnotemark[2]\\
0.05&0.1066(1)\footnotemark[2]\\
0.1&0.111~158(1)\\
$\frac{1}{3}$&0.133~519(3)\footnotemark[3]\\
0.5&0.152~4755(5)\\
0.9&0.2334(1)\footnotemark[2]\\
0.95&0.258~112(2)\\
0.99&0.2968(1)\footnotemark[2]\\
1&$\frac{1}{3}\footnotemark[4]$
\end{tabular}
\end{ruledtabular}
\footnotetext[1]{From Ref.~\cite{PP2007}.}
\footnotetext[2]{detailed analysis not shown in the paper.}
\footnotetext[3]{From Ref.~\cite{PP2006}.}
\footnotetext[4]{Mean field critical point. See Sec. \ref{Sec:MFcross}.}
\end{table}
\section{\label{Sec:cross}Crossover}
To get a hint to the crossover behavior occurring at two limiting cases
$d=0$ and $d\rightarrow 1$, we found critical points for a wide
range of $d$, which are summarized in Table~\ref{Table:pc}.
Because the estimate of the critical points within
error $10^{-4}$ is relatively easy with the present
computing power, we just present the resulting critical points without
showing the details. 

\begin{figure}[t]
\includegraphics[width=\columnwidth]{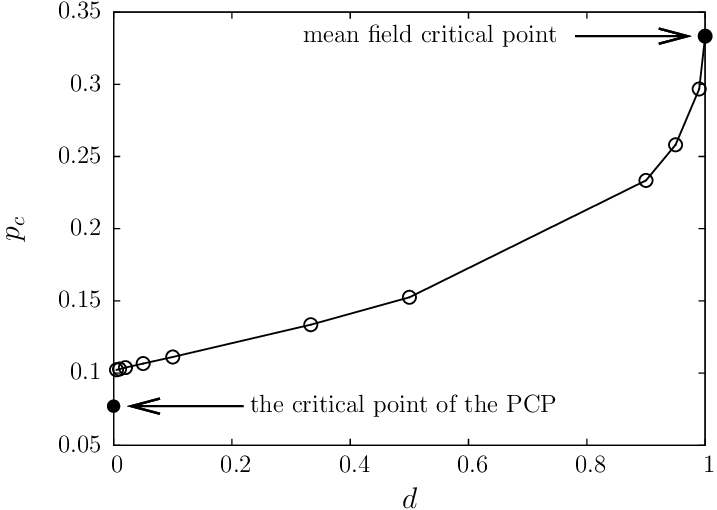}
\caption{\label{Fig:pb} Plot of $p_c$ vs $d$ for the PCPD. The critical point of 
the PCP without diffusion and the mean field critical point are indicated by respective arrows.}
\end{figure}

The phase boundary of the PCPD in the $d$-$p$ plane shown in
Fig.~\ref{Fig:pb} has two salient features 
at the two end points, $d=0$ and $d=1$. The phase boundary is discontinuous at $d=0$ and  
the phase boundary approaches the mean field critical point with infinite slope as 
$d\rightarrow 1$. Each singular behavior
signifies a crossover; crossover from the PCP to the PCPD 
at $d=0$ and that from the mean field
PCPD to the one dimensional PCPD at $d=1$. In this section, we will investigate these two kinds of
crossover behavior one by one and find the corresponding crossover exponents.

\subsection{\label{Sec:pcp}From the PCP to the PCPD}
\begin{figure}[t]
\includegraphics[width=\columnwidth]{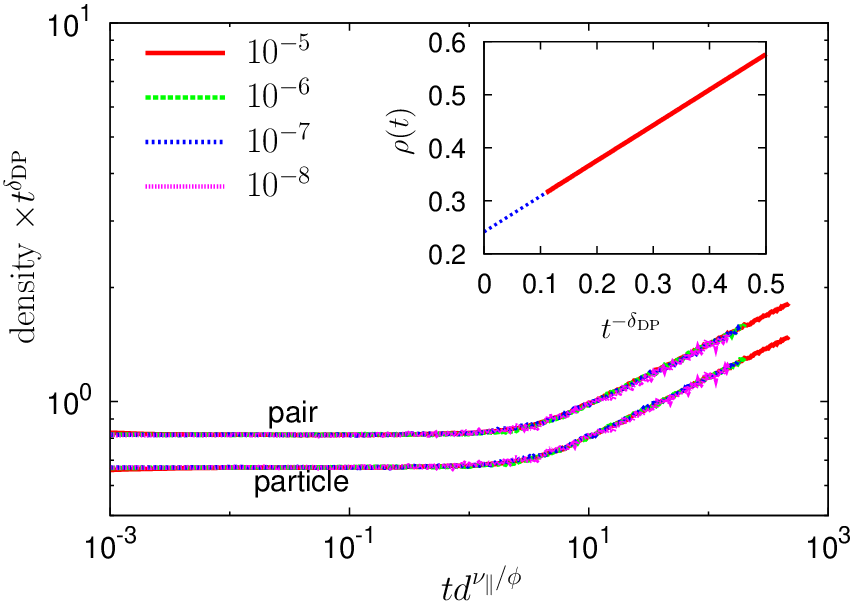}
\caption{\label{Fig:pcpcross} (Color online) Log-log plots of $\rho_p t^{\delta_\text{DP}}$ 
vs $t d^{\nu_\|/\phi}$ (top curves) and $(\rho - \rho_s) t^{\delta_\text{DP}}$ vs
$t d^{\nu_\|/\phi}$ (bottom curves) for $d = 10^{-5}$, $10^{-6}$, $10^{-7}$, and $10^{-8}$. 
$\rho_s$ is the steady state density of particles for $d=0$, which is estimated as $\approx 0.2414$.
$\delta_\text{DP} \simeq 0.1595$ and $\nu_\| \simeq 1.732$ are the critical exponents of the DP.
In the scaling collapse, we use $\phi= 2.6$. Inset: Plot of $\rho(t)$ vs $t^{-\delta_\text{DP}}$
at the PCP critical point. The extrapolation gives $\rho_s= 0.2414(3)$.
}
\end{figure}

The discontinuity of the phase boundary at $d=0$ 
can be understood as follows. In this discussion, all exponents are of the DP class.
Consider the system at $p = p_c(0)+\Delta p $ with $0< \Delta p \ll p_c(0)$, where 
$p_c(0)$ is the critical point of the PCP. If $d = 0$, the system is in the absorbing phase
and the pair density decays exponentially if time $t$ exceeds the
relaxation time $\xi_t \sim \Delta p^{-\nu_\|}$.
On the other hand, the particle density should approach a certain nonzero value $\rho_s(p)$.
If $0<d \xi_t \ll \rho_s(p)^{-2}$, it is unlikely for isolated particles to meet each other purely by diffusion when $t$ is smaller than $\xi_t$.
Hence, effectively, the initial PCP dynamics is almost decoupled with diffusion before $t=\xi_t$ and
only when $t$ exceeds $\xi_t$ pairs formed by diffusion can appear. 
Since we are considering an infinite system, the pair density is nonzero for any finite $t$ although
it can be extremely small. As soon as pairs
appear by diffusion, the so-called defect dynamics of the PCP begins.
Since the probability that two consecutive sites are occupied by diffusion is roughly $\rho_s(p)^2$,
the mean distance between two pairs formed by diffusion should be $1/\rho_s(p)^2$. 
Let $P(\xi_x)$ be the probability that the
defect dynamics continues until the cluster size becomes $\xi_x \sim \Delta p^{-\nu_\perp}$.
$P(\xi_x)$ should be the same order as the survival probability that the defect dynamics 
continues until $\xi_t$, which for small $\Delta p$ becomes $\xi_t^{-\beta/\nu_\|} \sim \Delta p^\beta$.
This is because the scaling form of the survival probability is $t^{-\beta/\nu_\|} f(t/\xi_t)$, where
 $f(x)$ is a scaling function with finite value of $f(1)$.
Then the mean distance between the starting points of two successfully spreading clusters
should be $\sim 1/[\rho_s(p)^2 P(\xi_x)]$.
If $\xi_x \gg 1 /[\rho_s(p)^2 P(\xi_x)] $ or $\rho_s(p)^2 \xi_x P(\xi_x) \gg 1$, 
a merger of two spreading clusters into a single cluster happens frequently and the system should survive 
indefinitely.  Hence, the condition that the system dynamics continue indefinitely by any small but finite 
$d$ is $\rho_s(p)^2 \xi_x P(\xi_x) \gg 1$ or $\rho_s(p)^2 (\Delta p)^{-\nu_\perp + \beta} \gg 1$.
Since $\nu_\perp > \beta$, there should be a finite range of $\Delta p$ which satisfies
the above criterion. Thus, the phase boundary should be discontinuous at $d=0$.
Using $\rho_s(p)$ at the critical point of the PCP (see the inset of Fig.~\ref{Fig:pcpcross})
and the DP exponents $\beta \approx 0.27$, $\nu_\perp \approx 1.09$, the validity of the above criterion 
gives $\Delta p < 0.03$ which is comparable to the numerical result.

Due to the discontinuity, the phase boundary does not give any information about the crossover 
exponent $\phi$. Rather, we find $\phi$ by data collapse using the following scaling ansatz,
\begin{eqnarray}
\rho(p,d;t) &=& \rho_s(p) + t^{-\delta_\text{DP}} \Psi_\rho\left (t | p-p_c(0)|^{\nu_\|}, t d^{\nu_\|/\phi} \right ),
\nonumber\\
\rho_p(p,d;t) &=& t^{-\delta_\text{DP}} \Psi_p\left (t | p-p_c(0)|^{\nu_\|}, t d^{\nu_\|/\phi} \right ),
\label{Eq:PCPcross}
\end{eqnarray}
where $\rho_s(p)$ is the steady state particle density at $d=0$, $\phi$ is the crossover exponent, 
$\Psi_\rho$, $\Psi_p$ are scaling functions, and $\delta_\text{DP} \approx 0.1595$ and 
$\nu_\| \approx 1.732$ are the critical exponents of the DP class.
We observe the best collapse when we use $\phi=2.6$ as shown in Fig.~\ref{Fig:pcpcross}.
Thus, we conclude $\phi = 2.6(1)$.
Note that this crossover exponent is different from that of the crossover from 
the DP class with infinitely many absorbing states to the DP class with finite number of
absorbing states~\cite{PP2007}.

The existence of the nontrivial crossover behavior at $d=0$ has
nothing to do with the change of universality classes.
In fact, this crossover originates from
the drastic decrease of the volume of absorbing states in the configuration
space~\cite{PP2007}.
However, the discontinuity at $d=0$ in the phase boundary raises a criticism 
on the Hinrichsen's argument explained in Sec.~\ref{Sec:intro}.
This criticism starts from numerical observation that 
diffusion makes the system more active as the system
at $p=p_c(0)$ with nonzero $d$ is in the active phase.
Now consider the system at $p=p_c(0)$ and $0< d \ll 1$. In this case, 
clusters spreads even faster than the critical spreading in the
long time limit.
Repeating Hinrichsen's argument, one can conclude that 
diffusion of isolated particles is negligible in the long time limit
and, in turn, isolated particles can join a cluster 
by the spreading of clusters not by their own diffusion, 
which is the crucial feature of immobile isolated particles in the PCP.
If this were the case, the critical point should approach $p_c(0)$ as
$d\rightarrow 0$ and the phase boundary should be continuous at $d=0$ 
just as the inhibitory route in Ref.~\cite{PP2007}. 
Obviously, this is contradictory to the numerical result. 
Also note that the origin of the discontinuity is the active role
of isolated particles, which is completely neglected in the 
Hinrichsen's argument. In other words, one cannot deduce a right conclusion 
by simply comparing the speed of spreading clusters with that of diffusion.
Hence, we cannot neglect the effect of pure diffusion
and there should be a strong correlation between diffusion and the critical cluster spreading, which can mediate the change of the universality class.

\subsection{\label{Sec:MFcross}From the mean field PCPD to the PCPD}
The mean field equation for the PCPD is obtained by setting $\rho_p(t) =\rho(t)^2$
and $\rho_t(t) = \rho(t)^3$ in
Eq.~\eqref{Eq:EE}, which gives
\begin{equation}
\frac{d\rho}{d \tau} = (1-3 p) \rho^2 - (1-p) \rho^3,
\label{Eq:MFE}
\end{equation}
where $\tau = (1-d)t$. The critical point of the mean field theory
is $p_0 = \frac{1}{3}$ at which the density decays as
\begin{equation}
\rho_c(\tau) = \frac{\rho_0}{\left ( 1 + 4 \rho_0^2 \tau/3 \right )^{1/2}} \sim \tau^{-1/2},
\label{Eq:MF}
\end{equation}
where $\rho_0$ is the initial density (we will set $\rho_0 = 1$).
When $3 |p_0-p| \rho_c^2 \ll (1-p) \rho_c^3$, $\rho(t)$ is indistinguishable
from $\rho_c(t)$ and clear deviation from Eq.~\eqref{Eq:MF}
is observable when $3 |p_0-p| \rho_c^2 \gg (1-p) \rho_c^3$, 
or $|p-p_0|\tau^{-1} \gg  \tau^{-3/2}$. 
In other words, the critical density decay is observable when 
$\tau \ll |p-p_0|^{-2}$ and the off-critical 
behavior dominates when $\tau \gg |p-p_0|^{-2}$. Hence, the critical exponent $\nu_\|$ for the mean field 
theory is 2. 

The relation between the mean field equation and the PCPD under $d\rightarrow 1$ limit can be
understood as follows (similar argument is also found in Ref.~\cite{P2009}).
Under the limit $d\rightarrow 1$ with $\tau$ kept finite, any correlation generated by reaction dynamics 
will be removed by diffusion immediately. Since the mean field theory assumes no correlation for all time 
$\tau$, Eq.~\eqref{Eq:MFE} becomes exact in this limit.
Even for finite $1-d$, the mean field equation is an accurate approximation if 
the relaxation time of diffusion of a randomly chosen region is much smaller 
than the time between two consecutive reaction dynamics in the same region. 

\begin{figure}[t]
\includegraphics[width=\columnwidth]{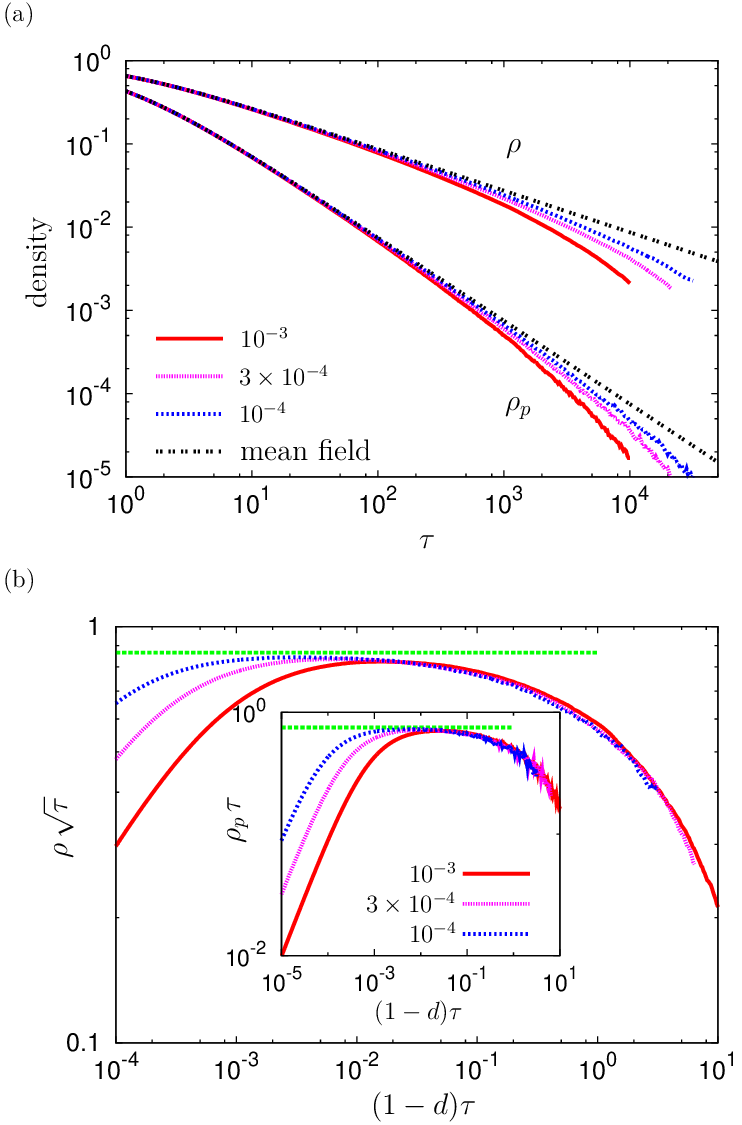}
\caption{\label{Fig:MFcross} (Color online) (a) Plots of particle density $\rho$ 
and pair density $\rho_p$  as functions of $\tau = (1-d) t$ for
$1-d = 10^{-3}$, $3\times 10^{-4}$, $10^{-4}$, and the mean field theory (bottom to top for each group). 
(b) Scaling collapse plot of $\rho \sqrt{\tau}$ vs $(1-d) \tau$ at $p=p_0$ for 
same $d$'s in (a). For comparison, $\Phi_\rho(0,0) = \sqrt{3}/2$, is drawn as a line segment. 
Inset: Scaling collapse plots of
$\rho_p \tau$ vs $(1-d) \tau$ for the same values of $d$ For comparison
$\Phi_p(0,0) = \frac{3}{4}$ is drawn as a line segment.
}
\end{figure}
To find the criterion for the validity of the mean field theory for small but 
finite $1-d$, consider the mean field dynamics at $p=p_0$. According to the mean field solution, 
the mean distance $\ell(\tau)$ at time $\tau$ between two nearest particles is 
$\ell(\tau) = \rho^{-1} \sim \tau^{1/2}$.
Now consider a region of size $O(\ell(\tau))$ at time $\tau$ and assume that two consecutive reaction events
occur at $\tau$ and $\tau + \Delta \tau$ in this region. 
We also assume that during $\Delta \tau$, this region is not correlated with outside of this region.
Since the number of particles is finite in this region, $\Delta \tau$ should be $O(1)$; recall that
$\tau$ is rescaled time.
Since the diffusion constant in rescaled time $\tau$ is $d/(1-d) \approx 1/(1-d)$, the relaxation time
of diffusion of this region is $O(\ell(\tau)^2) (1-d) \sim (1-d) \tau$. 
If $(1-d) \tau \ll \Delta \tau \sim 1$, two consecutive reaction events are uncorrelated and the
mean field theory becomes accurate. 
On the other hand, if $(1-d) \tau \gg 1$, two consecutive reaction events become correlated and
the mean field theory fail to describe the system correctly. 
Hence, the crossover from the mean field PCPD to the PCPD occurs when  $\tau \sim (1-d)^{-1}$
and $(1-d) \tau$ becomes a proper scaling parameter.

According to the above argument, the particle density in the asymptotic regime should take the scaling form
\begin{equation}
\rho(p,d,\tau) = \tau^{-1/2} \Phi_\rho(\tau |p-p_0|^2, \tau (1-d)),
\end{equation}
where $\tau = (1-d)t$ as above and $\Phi_\rho$ is a scaling function.
We also conjecture the scaling form of the pair density as
\begin{equation}
\rho_p(p,d,\tau) = \tau^{-1} \Phi_p(\tau |p-p_0|^2, \tau (1-d)),
\end{equation}
where $\Phi_p$ is another scaling function.
From the mean field theory, $\Phi_\rho(0,0) = \sqrt{3}/2$ and $\Phi_p(0,0) = 3/4$.
Hence, if $1-d \ll 1$ and $p = p_0$, plots of $t^{1/2} \rho(p_0,d,\tau)$ against $(1-d) \tau$
should collapse onto a single curve for sufficiently large $(1-d) \tau$. Furthermore,
the phase boundary for $1-d \ll 1$ should approach the mean field critical point as
\begin{equation}
|p_c(d) - p_0 | \sim (1-d)^{1/2},
\end{equation}
where $p_c(d)$ is the critical point of the PCPD for given $d<1$.
Hence, the crossover exponent is $\phi = 2$.

To confirm the above argument, we simulated the PCPD for $1-d = 10^{-3}$, $3 \times 10^{-4}$, 
and $10^{-4}$ at $p = \frac{1}{3}$ with the system size $L = 2^{22}$.
In Fig.~\ref{Fig:MFcross}(a), we depict $\rho(\tau)$ vs $\tau$ and $\rho_p(\tau)$ vs $\tau$ on 
a double-logarithmic scale. For comparison, the mean field solution is also depicted. As argued, 
the regime where mean field theory is 
accurate becomes larger as $d$ gets closer to 1.
In Fig.~\ref{Fig:MFcross}(b), one can see a scaling collapse plot of $\rho \sqrt{\tau}$ vs
$(1-d) \tau$ as well as $\rho_p \tau$ vs $(1-d) \tau$, which affirms that the crossover exponent is $2$.

\section{\label{Sec:sum}Summary}
To sum up, we numerically studied the critical density decay of the pair contact process with diffusion (PCPD)
and estimated the critical decay exponent by investigating effective exponents after finding 
corrections to scaling for various diffusion strength. For small diffusion probability ($d \le 0.5$), 
we found that the corrections-to-scaling term asymptotically behaves as $t^{-0.15}$ and for large diffusion 
probability ($d =0.95$) the corrections-to-scaling term decays as $t^{-0.5}$ which is 
weaker than that of the case with small $d$. All the same, the analysis of the effective exponents
for any $d$ with the corresponding corrections-to-scaling term showed that
the critical decay exponent is $\delta = 0.173(3)$.
Although this value is quite close to that of the directed percolation (DP) universality class
which is 0.1595, the systematic deviation of $\delta$ for the PCPD from the DP value for any $d$
suggests that the PCPD does not belong to the DP class and forms an independent universality
class.

We also studied the crossover from the pair contact process (PCP) without diffusion to
the PCPD which occurs around $d=0$ and from the mean field theory (MFT) to the PCPD which 
happens around $d=1$. We found that the crossover at $d=0$ is characterized
by the discontinuity of the phase boundary and that the crossover exponent is
$\phi=2.6(1)$. We showed that applying the Hinrichsen's argument~\cite{H2006} which supports the 
DP hypothesis to this crossover leads to a contradictory conclusion to 
the discontinuity of the phase boundary at $d=0$. 
The crossover from the MFT to the PCPD, occurring at $d=1$, is described
by the crossover exponent $\phi = 2$, which was argued to be exact.

\begin{acknowledgments}
This work was supported by the Basic Science Research Program
through the National Research Foundation of Korea
(NRF) funded by the Ministry of Education, Science
and Technology (Grant No. 2011-0014680);
and by the Catholic University of Korea, Research Fund, 2013.
The author acknowledges the hospitality of the Institut f\"ur Theoretische 
Physik, Universit\"at zu K\"oln, Germany and support under a German Research Foundation (DFG) grant within 
SFB 680 {\it Molecular Basis of Evolutionary Innovations} 
during the final stage of this work. The computation was 
supported by Universit\"at zu K\"oln.
The author also would like to thank the APCTP for its hospitality.
\end{acknowledgments}
\bibliographystyle{apsrev}
\bibliography{Park}

\begin{thebibliography}{29}
\expandafter\ifx\csname natexlab\endcsname\relax\def\natexlab#1{#1}\fi
\expandafter\ifx\csname bibnamefont\endcsname\relax
  \def\bibnamefont#1{#1}\fi
\expandafter\ifx\csname bibfnamefont\endcsname\relax
  \def\bibfnamefont#1{#1}\fi
\expandafter\ifx\csname citenamefont\endcsname\relax
  \def\citenamefont#1{#1}\fi
\expandafter\ifx\csname url\endcsname\relax
  \def\url#1{\texttt{#1}}\fi
\expandafter\ifx\csname urlprefix\endcsname\relax\def\urlprefix{URL }\fi
\providecommand{\bibinfo}[2]{#2}
\providecommand{\eprint}[2][]{\url{#2}}

\bibitem[{\citenamefont{Grassberger}(1982)}]{G1982}
\bibinfo{author}{\bibfnamefont{P.}~\bibnamefont{Grassberger}},
  \bibinfo{journal}{Z. Phys. B} \textbf{\bibinfo{volume}{47}},
  \bibinfo{pages}{365} (\bibinfo{year}{1982}).

\bibitem[{\citenamefont{Howard and T{\"a}uber}(1997)}]{HT1997}
\bibinfo{author}{\bibfnamefont{M.~J.} \bibnamefont{Howard}} \bibnamefont{and}
  \bibinfo{author}{\bibfnamefont{U.~C.} \bibnamefont{T{\"a}uber}},
  \bibinfo{journal}{J. Phys. A} \textbf{\bibinfo{volume}{30}},
  \bibinfo{pages}{7721} (\bibinfo{year}{1997}).

\bibitem[{\citenamefont{\'Odor}(2000)}]{O2000}
\bibinfo{author}{\bibfnamefont{G.}~\bibnamefont{\'Odor}},
  \bibinfo{journal}{Phys. Rev. E} \textbf{\bibinfo{volume}{62}},
  \bibinfo{pages}{R3027} (\bibinfo{year}{2000}).

\bibitem[{\citenamefont{Carlon et~al.}(2001)\citenamefont{Carlon, Henkel, and
  Schollw\"ock}}]{CHS2001}
\bibinfo{author}{\bibfnamefont{E.}~\bibnamefont{Carlon}},
  \bibinfo{author}{\bibfnamefont{M.}~\bibnamefont{Henkel}}, \bibnamefont{and}
  \bibinfo{author}{\bibfnamefont{U.}~\bibnamefont{Schollw\"ock}},
  \bibinfo{journal}{Phys. Rev. E} \textbf{\bibinfo{volume}{63}},
  \bibinfo{pages}{036101} (\bibinfo{year}{2001}).

\bibitem[{\citenamefont{Hinrichsen}(2001)}]{H2001}
\bibinfo{author}{\bibfnamefont{H.}~\bibnamefont{Hinrichsen}},
  \bibinfo{journal}{Phys. Rev. E} \textbf{\bibinfo{volume}{63}},
  \bibinfo{pages}{036102} (\bibinfo{year}{2001}).

\bibitem[{\citenamefont{Park and Kim}(2002)}]{PK2002}
\bibinfo{author}{\bibfnamefont{K.}~\bibnamefont{Park}} \bibnamefont{and}
  \bibinfo{author}{\bibfnamefont{I.-M.} \bibnamefont{Kim}},
  \bibinfo{journal}{Phys. Rev. E} \textbf{\bibinfo{volume}{66}},
  \bibinfo{pages}{027106} (\bibinfo{year}{2002}).

\bibitem[{\citenamefont{Barkema and Carlon}(2003)}]{BC2003}
\bibinfo{author}{\bibfnamefont{G.~T.} \bibnamefont{Barkema}} \bibnamefont{and}
  \bibinfo{author}{\bibfnamefont{E.}~\bibnamefont{Carlon}},
  \bibinfo{journal}{Phys. Rev. E} \textbf{\bibinfo{volume}{68}},
  \bibinfo{pages}{036113} (\bibinfo{year}{2003}).

\bibitem[{\citenamefont{Kockelkoren and Chat{\'e}}(2003)}]{KC2003}
\bibinfo{author}{\bibfnamefont{J.}~\bibnamefont{Kockelkoren}} \bibnamefont{and}
  \bibinfo{author}{\bibfnamefont{H.}~\bibnamefont{Chat{\'e}}},
  \bibinfo{journal}{Phys. Rev. Lett.} \textbf{\bibinfo{volume}{90}},
  \bibinfo{pages}{125701} (\bibinfo{year}{2003}).

\bibitem[{\citenamefont{Janssen et~al.}(2004)\citenamefont{Janssen, van
  Wijland, Deloubriere, and T{\"a}uber}}]{JvWDT2004}
\bibinfo{author}{\bibfnamefont{H.-K.} \bibnamefont{Janssen}},
  \bibinfo{author}{\bibfnamefont{F.}~\bibnamefont{van Wijland}},
  \bibinfo{author}{\bibfnamefont{O.}~\bibnamefont{Deloubriere}},
  \bibnamefont{and} \bibinfo{author}{\bibfnamefont{U.~C.}
  \bibnamefont{T{\"a}uber}}, \bibinfo{journal}{Phys. Rev. E}
  \textbf{\bibinfo{volume}{70}}, \bibinfo{pages}{056114}
  (\bibinfo{year}{2004}).

\bibitem[{\citenamefont{Noh and Park}(2004)}]{NP2004}
\bibinfo{author}{\bibfnamefont{J.~D.} \bibnamefont{Noh}} \bibnamefont{and}
  \bibinfo{author}{\bibfnamefont{H.}~\bibnamefont{Park}},
  \bibinfo{journal}{Phys. Rev. E} \textbf{\bibinfo{volume}{69}},
  \bibinfo{pages}{016122} (\bibinfo{year}{2004}).

\bibitem[{\citenamefont{Park and Park}(2005{\natexlab{a}})}]{PHP2005a}
\bibinfo{author}{\bibfnamefont{S.-C.} \bibnamefont{Park}} \bibnamefont{and}
  \bibinfo{author}{\bibfnamefont{H.}~\bibnamefont{Park}},
  \bibinfo{journal}{Phys. Rev. Lett.} \textbf{\bibinfo{volume}{94}},
  \bibinfo{pages}{065701} (\bibinfo{year}{2005}{\natexlab{a}}).

\bibitem[{\citenamefont{Park and Park}(2005{\natexlab{b}})}]{PHP2005b}
\bibinfo{author}{\bibfnamefont{S.-C.} \bibnamefont{Park}} \bibnamefont{and}
  \bibinfo{author}{\bibfnamefont{H.}~\bibnamefont{Park}},
  \bibinfo{journal}{Phys. Rev. E} \textbf{\bibinfo{volume}{71}},
  \bibinfo{pages}{016137} (\bibinfo{year}{2005}{\natexlab{b}}).

\bibitem[{\citenamefont{de~Oliveira and Dickman}(2006)}]{dOD2006}
\bibinfo{author}{\bibfnamefont{M.~M.} \bibnamefont{de~Oliveira}}
  \bibnamefont{and} \bibinfo{author}{\bibfnamefont{R.}~\bibnamefont{Dickman}},
  \bibinfo{journal}{Phys. Rev. E} \textbf{\bibinfo{volume}{74}},
  \bibinfo{pages}{011124} (\bibinfo{year}{2006}).

\bibitem[{\citenamefont{Park and Park}(2006)}]{PP2006}
\bibinfo{author}{\bibfnamefont{S.-C.} \bibnamefont{Park}} \bibnamefont{and}
  \bibinfo{author}{\bibfnamefont{H.}~\bibnamefont{Park}},
  \bibinfo{journal}{Phys. Rev. E} \textbf{\bibinfo{volume}{73}},
  \bibinfo{pages}{025105} (\bibinfo{year}{2006}).

\bibitem[{\citenamefont{Kwon and Kim}(2007)}]{KK2007}
\bibinfo{author}{\bibfnamefont{S.}~\bibnamefont{Kwon}} \bibnamefont{and}
  \bibinfo{author}{\bibfnamefont{Y.}~\bibnamefont{Kim}},
  \bibinfo{journal}{Phys. Rev. E} \textbf{\bibinfo{volume}{75}},
  \bibinfo{pages}{042103} (\bibinfo{year}{2007}).

\bibitem[{\citenamefont{Smallenburg and Barkema}(2008)}]{SB2008}
\bibinfo{author}{\bibfnamefont{F.}~\bibnamefont{Smallenburg}} \bibnamefont{and}
  \bibinfo{author}{\bibfnamefont{G.~T.} \bibnamefont{Barkema}},
  \bibinfo{journal}{Phys. Rev. E} \textbf{\bibinfo{volume}{78}},
  \bibinfo{pages}{031129} (\bibinfo{year}{2008}).

\bibitem[{\citenamefont{Park and Park}(2009)}]{PP2009}
\bibinfo{author}{\bibfnamefont{S.-C.} \bibnamefont{Park}} \bibnamefont{and}
  \bibinfo{author}{\bibfnamefont{H.}~\bibnamefont{Park}},
  \bibinfo{journal}{Phys. Rev. E} \textbf{\bibinfo{volume}{79}},
  \bibinfo{pages}{051130} (\bibinfo{year}{2009}).

\bibitem[{\citenamefont{Schram and Barkema}(2012)}]{SB2012}
\bibinfo{author}{\bibfnamefont{R.~D.} \bibnamefont{Schram}} \bibnamefont{and}
  \bibinfo{author}{\bibfnamefont{G.~T.} \bibnamefont{Barkema}},
  \bibinfo{journal}{J. Stat. Mech.:Theory Exp.}
  \textbf{\bibinfo{volume}{(2012)}}, \bibinfo{pages}{P03009}.

\bibitem[{\citenamefont{Gredat et~al.}(2014)\citenamefont{Gredat, Chat\'e,
  Delamotte, and Dornic}}]{GCDD2014}
\bibinfo{author}{\bibfnamefont{D.}~\bibnamefont{Gredat}},
  \bibinfo{author}{\bibfnamefont{H.}~\bibnamefont{Chat\'e}},
  \bibinfo{author}{\bibfnamefont{B.}~\bibnamefont{Delamotte}},
  \bibnamefont{and} \bibinfo{author}{\bibfnamefont{I.}~\bibnamefont{Dornic}},
  \bibinfo{journal}{Phys. Rev. E} \textbf{\bibinfo{volume}{89}},
  \bibinfo{pages}{010102} (\bibinfo{year}{2014}).

\bibitem[{\citenamefont{\'{O}dor et~al.}(2002)\citenamefont{\'{O}dor, Marques,
  and Santos}}]{OMS2002}
\bibinfo{author}{\bibfnamefont{G.}~\bibnamefont{\'{O}dor}},
  \bibinfo{author}{\bibfnamefont{M.~C.} \bibnamefont{Marques}},
  \bibnamefont{and} \bibinfo{author}{\bibfnamefont{M.~A.}
  \bibnamefont{Santos}}, \bibinfo{journal}{Phys. Rev. E}
  \textbf{\bibinfo{volume}{65}}, \bibinfo{pages}{056113}
  (\bibinfo{year}{2002}).

\bibitem[{\citenamefont{Henkel and Hinrichsen}(2004)}]{HH2004}
\bibinfo{author}{\bibfnamefont{M.}~\bibnamefont{Henkel}} \bibnamefont{and}
  \bibinfo{author}{\bibfnamefont{H.}~\bibnamefont{Hinrichsen}},
  \bibinfo{journal}{J. Phys. A} \textbf{\bibinfo{volume}{37}},
  \bibinfo{pages}{R117} (\bibinfo{year}{2004}).

\bibitem[{\citenamefont{Hinrichsen}(2006)}]{H2006}
\bibinfo{author}{\bibfnamefont{H.}~\bibnamefont{Hinrichsen}},
  \bibinfo{journal}{Physica A} \textbf{\bibinfo{volume}{361}},
  \bibinfo{pages}{457} (\bibinfo{year}{2006}).

\bibitem[{\citenamefont{Park and Park}(2008)}]{PP2008EPJB}
\bibinfo{author}{\bibfnamefont{S.-C.} \bibnamefont{Park}} \bibnamefont{and}
  \bibinfo{author}{\bibfnamefont{H.}~\bibnamefont{Park}},
  \bibinfo{journal}{Eur. Phys. J. B} \textbf{\bibinfo{volume}{64}},
  \bibinfo{pages}{415} (\bibinfo{year}{2008}).

\bibitem[{\citenamefont{Janssen}(1981)}]{J1981}
\bibinfo{author}{\bibfnamefont{H.-K.} \bibnamefont{Janssen}},
  \bibinfo{journal}{Z. Phys. B} \textbf{\bibinfo{volume}{42}},
  \bibinfo{pages}{151} (\bibinfo{year}{1981}).

\bibitem[{\citenamefont{Park}(2013)}]{P2013}
\bibinfo{author}{\bibfnamefont{S.-C.} \bibnamefont{Park}}, \bibinfo{journal}{J.
  Korean Phys. Soc.} \textbf{\bibinfo{volume}{62}}, \bibinfo{pages}{469}
  (\bibinfo{year}{2013}).

\bibitem[{\citenamefont{Jensen}(1993)}]{J1993}
\bibinfo{author}{\bibfnamefont{I.}~\bibnamefont{Jensen}},
  \bibinfo{journal}{Phys. Rev. Lett.} \textbf{\bibinfo{volume}{70}},
  \bibinfo{pages}{1465} (\bibinfo{year}{1993}).

\bibitem[{\citenamefont{\'Odor}(2003)}]{O2003}
\bibinfo{author}{\bibfnamefont{G.}~\bibnamefont{\'Odor}},
  \bibinfo{journal}{Phys. Rev. E} \textbf{\bibinfo{volume}{67}},
  \bibinfo{pages}{016111} (\bibinfo{year}{2003}).

\bibitem[{\citenamefont{Park and Park}(2007)}]{PP2007}
\bibinfo{author}{\bibfnamefont{S.-C.} \bibnamefont{Park}} \bibnamefont{and}
  \bibinfo{author}{\bibfnamefont{H.}~\bibnamefont{Park}},
  \bibinfo{journal}{Phys. Rev. E} \textbf{\bibinfo{volume}{76}},
  \bibinfo{pages}{051123} (\bibinfo{year}{2007}).

\bibitem[{\citenamefont{Park}(2009)}]{P2009}
\bibinfo{author}{\bibfnamefont{S.-C.} \bibnamefont{Park}},
  \bibinfo{journal}{Phys. Rev. E} \textbf{\bibinfo{volume}{80}},
  \bibinfo{pages}{061103} (\bibinfo{year}{2009}).

\end{thebibliography}
\end{document}